\documentclass[12pt]{iopart}

\usepackage{graphicx}

\def \Pr124{PrBa$_2$Cu$_4$O$_8$}

\begin{document}

\title[Fragile three-dimensionality in the quasi-one-dimensional
cuprate \Pr124
]{Fragile three-dimensionality in the quasi-one-dimensional
cuprate \Pr124}

\author{A Narduzzo$^1$, A Enayati-Rad$^1$, P J Heard$^2$, S L Kearns$^3$,
S Horii$^4$, F F Balakirev$^5$ and N E Hussey$^1$}

\address{$^1$ H. H. Wills Physics Laboratory, University of Bristol, Tyndall
Avenue, Bristol, BS8 1TL, UK}
\address{$^2$ Interface Analysis Centre, Oldbury House, 121 St. Michael's
Hill, Bristol, BS2 8BS, UK}
\address{$^3$ Department of Earth Sciences, University of Bristol, Wills
Memorial Building, Queen's Road, Bristol, BS8 1RJ, UK}
\address{$^4$ Department of Applied Chemistry, University of Tokyo,
7-3-1 Hongo, Bunkyo-ku, Tokyo, 113-8656, Japan}
\address{$^5$ National High Magnetic Field Laboratory, Los Alamos
National Laboratory, Los Alamos, New Mexico, 87545, USA}
\ead{n.e.hussey@bristol.ac.uk}

\begin{abstract}
In this article we report on the experimental realization of
dimensional crossover phenomena in the chain compound \Pr124 using
temperature, high magnetic fields and disorder as independent
tuning parameters. In purer crystals of \Pr124, a highly
anisotropic three-dimensional Fermi-liquid state develops at low
temperatures. This metallic state is extremely susceptible to
disorder however and localization rapidly sets in. We show,
through quantitative comparison of the relevant energy scales,
that this metal/insulator crossover occurs precisely when the
scattering rate within the chain exceeds the interchain hopping
rate(s), i.e. once carriers become confined to a single conducting
element.
\end{abstract}




\maketitle

\section{Introduction}
There is growing evidence that in compounds containing isolated
atomic~\cite{Kim96} or molecular~\cite{Claessen02} chains, the
conventional Fermi-liquid picture of electron-like quasiparticles
fails. How the corresponding one-dimensional
state~\cite{Tomonaga50, Lutt63} of decoupled spin and charge
excitations emerges however remains unresolved, prompting the
search for compounds whose electronic ground state can be tuned
progressively towards one-dimensionality. In metals on the
boundary of one-dimensionality, the so-called
quasi-one-dimensional (quasi-1D) conductors, the Fermi surface
takes the form of pair(s) of parallel corrugated sheets in the
plane normal to the conducting chain(s). Provided the two
orthogonal interchain hopping energies $2t_{\perp}$  (which
determine the size of corrugation in each direction) are larger
than other relevant perturbations, hopping between chains is
coherent and in the absence of charge ordering, a 3D Fermi-liquid
ground state is stabilized at low temperature. If this corrugation
is 'smeared out' and the chains become decoupled however, theory
predicts~\cite{Tomonaga50, Lutt63} that even weak interactions
will drive the system into the 1D Luttinger-liquid state with its
associated phenomenon of spin-charge separation. In order to
realize and study the crossover between these two extreme ground
states, it is necessary to identify materials where $2t_{\perp}$
in both directions is restricted, due to orbital overlap or
correlation effects, to energies attainable within the laboratory.

\Pr124 (Pr124) is the non-superconducting analogue of the
high-temperature superconductor YBa$_{2}$Cu$_{4}$O$_{8}$ (Y124).
Its crystal structure, shown in Fig. 1), consists of edge-sharing
double CuO chain networks (oriented along the crystallographic
{\it b}-axis) sandwiched between sets of CuO$_{2}$ bilayer
plaquettes. Substitution of Pr for Y between the bilayers
completely suppresses superconductivity (and mobility) within the
CuO$_{2}$ planes~\cite{Horii98} whilst preserving the metallicity
of the double chains~\cite{Tera96, Horii00}. This offers a unique
opportunity to study the charge dynamics of the cuprate chain in
isolation. In this article, we show that temperature, magnetic
fields and disorder can all induce a 3D to 1D crossover in the
electronic ground state of Pr124 under laboratory conditions.
Whilst dimensional crossover phenomena due to high
temperatures~\cite{Hussey98a, Moser98}, high magnetic
fields~\cite{Behnia95, Hussey98b, Hussey02} and even strong
correlation effects~\cite{Vesco98} have been well documented, to
our knowledge, this is the first experimental realization of {\it
disorder}-induced one-dimensionality in a three-dimensional
compound.

\section{Methods}

\begin{figure}[t]
\centering
\includegraphics[height=7.5cm,clip]{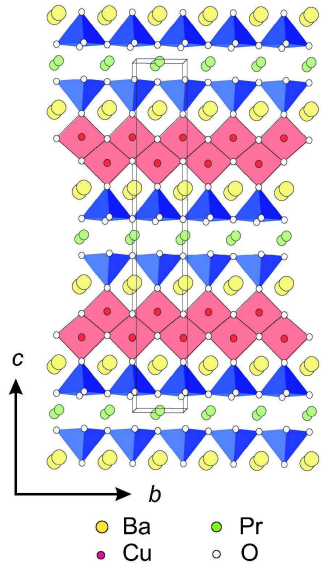}
\caption{Schematic of the crystallographic structure of Pr124. The
red squares represent the CuO double chain network oriented along
the $b$-axis. The unit cell is indicated by a thin solid line.}
\label{struct}
\end{figure}

\subsection{Crystal growth and characterisation}
Single crystals of \Pr124 were grown by a self-flux method in MgO
crucibles under high-pressure oxygen gas of 11atm~\cite{Horii00}.
The impurity content of three of the crystals used in the disorder
study was investigated by means of secondary-ion mass spectrometry
(SIMS) as well as electron probe micro-analysis (EPMA). SIMS
identified a number of trace impurity elements (including Fe but
not Ni or Co) but only three, Mg, Al and Sr, had abundances above
the detectability limit of our EPMA measurements (100ppm). Of
these, Mg was by far the most abundant, as expected through
contamination with the crucible walls.

\subsection{Zero- and pulsed-field resistivity measurements}
The resistivities were measured using a standard four-probe ac
lock-in technique. For the $\rho_{c}(T,B)$ measurements,
electrical contacts were mounted on the top surfaces of the
crystal whilst for the $\rho_{a}(T,B)$ measurements, they were
mounted on the corners. In each case, the other highly resistive
direction was shorted out to ensure uniaxial current flow. For
in-chain resistivity ($\rho_b$) measurements, the crystal
dimensions were recorded using a scanning electron microscope. In
addition, voltage contacts were placed so as to avoid
contamination from the two other highly resistive current
directions. An example of the voltage contact configuration for
$\rho_b$ measurements is shown in Fig. A.1) of the Appendix. Due
to the smallness of the samples and the finite width of the
voltage electrodes, the convention adopted in estimating their
distance plays an important role in the determination of the
absolute resistivity values. In this study, we have taken the
midpoint of the wire (diameter 25 $\mu$m) at the sample as the
marker for the electrode position. Uncertainties in our estimates
of the sample dimensions are 10-15\%. In previous
studies~\cite{Hussey02, McBrien02, Nakada01}, different markers
have been used. As discussed in the Appendix, we believe this
choice of marker is primarily responsible for the large
discrepancies in the absolute magnitudes of the $\rho_b$ values
reported in the literature.

The high field measurements were performed in the 65 Tesla pulsed
magnet at the NHMFL, Los Alamos, USA. A typical 100 ms long
magnetic field pulse is produced by discharging 1.6 MJ capacitor
bank through a reinforced copper alloy magnet coil. Resistance
versus magnetic field is recorded again using a standard high
frequency lock-in technique.



\section{Temperature and Magnetic Field Induced Dimensional Crossover Phenomena in Pr124}

It was shown previously~\cite{McBrien02} that in high-quality
Pr124 crystals, electrical resistivity at low $T$ is metallic in
all three orthogonal directions and varies approximately as
$T^{2}$, consistent with the development of a 3D Fermi-liquid
ground state. The resistive anisotropy at low $T$ however is
extremely large ($\rho_{a}:\rho_{b}:\rho_{c}(T=0)\sim
1000:1:3000$)~\cite{McBrien02}, with a similar anisotropy in the
ratio of the (squares of the) hopping energies. Moreover, whilst
$\rho_{b}(T)$ in the purest crystals remains metallic for all
$T<300$K, the interchain resistivities $\rho_{a}(T)$ and
$\rho_{c}(T)$ have maxima around $T=150$K above which their
behaviour becomes thermally activated. These maxima have been
interpreted either as a 3D to 1D crossover with increasing
temperature or as the emergence of a contribution to
$\rho_{a,c}(T)$ from the insulating CuO$_{2}$ planes.

Dimensional crossover phenomena can also be realized in high
magnetic fields, due to a field-induced real-space confinement of
the charge carriers~\cite{Gorkov84}. In the double-chain cuprate
Pr124, the Fermi surface consists of two pairs of corrugated
sheets extending normal to the reciprocal space axis ${\mathbf
k}_{b}$. Within a simple tight-binding picture, the $c$-axis
dispersion (for a single chain) is
$\mathcal{E}=-2t_{\perp}^{c}cos(k_{c}c)$. For $B\parallel a$, the
dominant Lorentz force $e[{\bf v}_{F}\times{\bf B}]=ev_{F}B{\bf
\hat{c}}=\hbar d{\bf k}_{c}/dt$ causes carriers to traverse the
Fermi sheet along $k_{c}$. The sinusoidal corrugation then gives
rise to an oscillatory component to the $c$-axis velocity
$v_{\perp}^{c}=\hbar^{-1}[\partial
\mathcal{E}_{\perp}^{c}/\partial
k_{z}]=[2t_{\perp}^{c}c/\hbar]sin(k_{c}c)$ and hence to a
real-space sinusoidal trajectory with amplitude
$A_{c}=2t_{\perp}^{c}/ev_{F}B$. Thus $A_{c}$ shrinks as $B$
increases until eventually at $B_{\rm
cr}^{c}=2t_{\perp}^{c}/ev_{F}c$, $A_{c}=c$ and the charge carriers
become confined to a single plane of coupled chains. Note that
$B_{\rm cr}^{c}$ is independent of $1/\tau$ and therefore
independent of temperature and impurity concentration, as verified
experimentally~\cite{Hussey02, Waku00}. Due to the quasi-1D nature
of the Fermi sheets in Pr124, a similar oscillatory component of
amplitude $A_{a}=2t_{\perp}^{a}/ev_{F}B$ will also be induced
along the $a$-axis for $B \parallel c$. In this case, the
crossover field is expected to occur at $B_{\rm
cr}^{a}=2t_{\perp}^{a}/ev_{F}a$.

Figure 2a) shows transverse field sweeps of the $c$-axis
resistivity $\rho_{c}(B)$ (transverse to both the current and the
conducting chain) at different fixed temperatures. The inset shows
a blow-up of the low-field region (enclosed by a solid rectangle
in Fig. 2a), minus the 40K data for clarity). Below a crossing
field $B_{\rm cr}^{c}$ (= 10(1) Tesla, shown by a dashed line),
$\rho_{c}(T)$ is metallic. Above $B_{\rm cr}^{c}$ however, the
trend is reversed, implying a metallic/non-metallic crossover in
the interchain resistivity as a function of field. (For more
details on the precise determination of $B_{\rm cr}^{c}$, please
refer to Ref.~\cite{Hussey02}). As shown in Fig. 2b), a similar
phenomenon occurs in the reciprocal configuration ($B\parallel c$,
$I\parallel a$), though here the corresponding crossover field is
extremely high, $B_{\rm cr}^{a}$ = 62(2) Tesla. This is due
principally to the fact the $a \sim c$/3 and therefore a much
larger field is required to confine the electrons along the
$a$-axis.

\begin{figure}[t]
\centering
\includegraphics[height=9.8cm, clip]{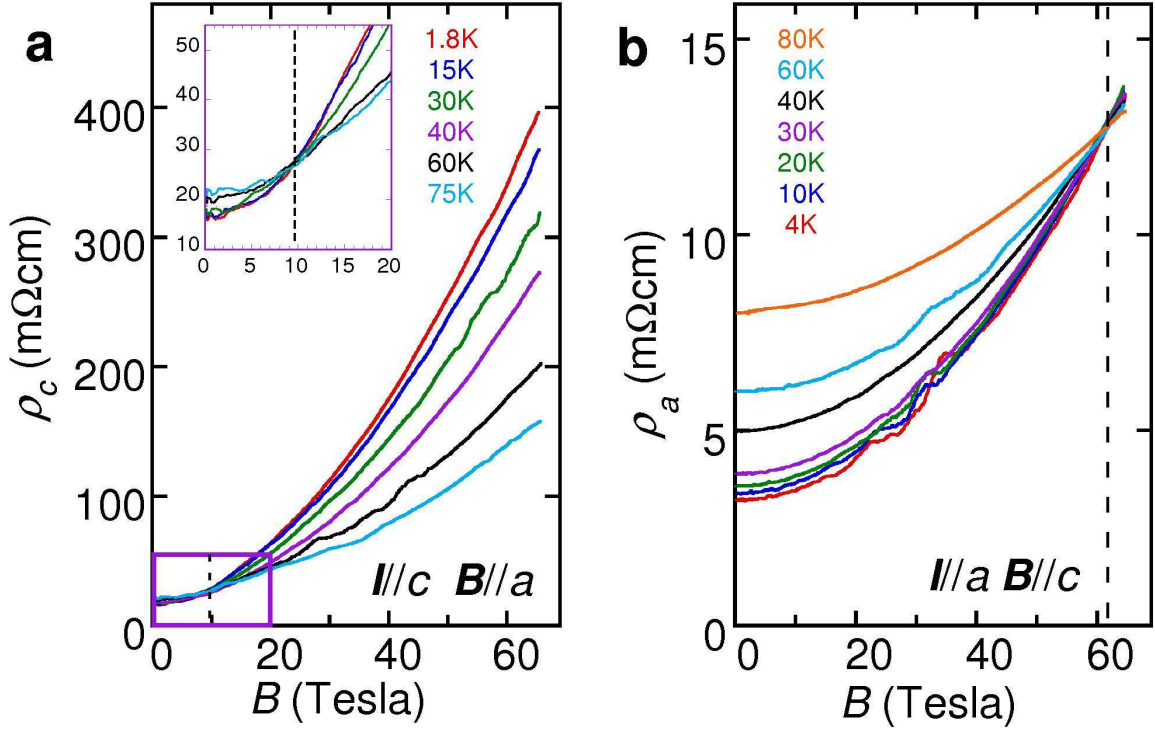}
\caption{{\bf a)} Transverse field sweeps of the $c$-axis
resistivity $\rho_{c}(B)$ (transverse to both the current and the
conducting chain) at different fixed temperatures. The inset shows
a blow-up of the low-field region (enclosed by a solid rectangle
in Fig. 1a), minus the 40K data for clarity). The dashed line
indicates the crossover field $B_{\rm cr}^{c}$ (see main text).
{\bf b)} Transverse field sweeps of the $a$-axis resistivity
$\rho_{a}(B)$. Again the dashed line indicates the crossover field
$B_{\rm cr}^{a}$.} \label{FIDC}
\end{figure}

From the crossover fields in the transverse $B$ sweeps shown in
Fig. 2a) - b), we obtain $2t_{\perp}^{c} \sim 45(5)$K and
$2t_{\perp}^{a} \sim 70(2)$K. We can now compare these values with
estimates of $2t_{\perp}$ from the $T$-dependent resistivity data.
Fig. 3a) and 3b) show $\rho_{c}(T)$ and $\rho_{a}(T)$ data
respectively for the same crystals that were used in the magnetic
field study. Typically, there are two energy scales that are used
as a measure of $2t_{\perp}$ in the interplane(chain)
resistivities of low-dimensional metals; the peak in
$\rho_{\perp}(T)$ at $T_{\rm max}^{a,c}$ is generally regarded as
an upper bound for 2$t_{\perp}$ whilst the deviation from the
low-$T$ quadratic resistivity gives a lower bound. In
Sr$_{2}$RuO$_{4}$ for example, the latter criterion has been shown
to give very consistent agreement with the value of $2t_{\perp}$
estimated for the most conducting band from quantum oscillation
experiments~\cite{Mac96}. The insets in Fig. 3a) and 3b) show
$\rho_{c,a}(T)$ versus $T^{2}$ below $T=70$K and $T=110$K
respectively. The arrows indicate the temperatures $T_{\rm
coh}^{a,c}$ at which $\rho_{a,c}(T)$ first deviates from $T^{2}$.
From these plots we find $T_{\rm coh}^{a} = 70(5)$K, $T_{\rm
max}^{a}$ = 130K, $T_{\rm coh}^{c} = 50(5)$K and $T_{\rm max}^{c}$
= 180K.  Note that the $T_{\rm coh}^{a,c}$ values are in excellent
agreement with the values for $2t_\perp$ determined from the two
sets of magnetic field measurements. From this we conclude that
$T_{\rm coh}^{a,c}$ defines the temperature at which $a$,$c$-axis
hopping first begins to lose coherence (i.e. when $k_{\rm B}T \sim
2t_{\perp}$). The peaks, in contrast, appear to represent the
temperature at which all interchain coherence is lost. In the
temperature range $T_{\rm coh}^{a,c}$ $\leq$ $T$ $\leq$ $T_{\rm
max}^{a,c}$ therefore, the chains are very weakly coupled and
metallicity is seen to disappear only gradually.

\begin{figure}[t]
\centering
\includegraphics[height=6.8cm, clip]{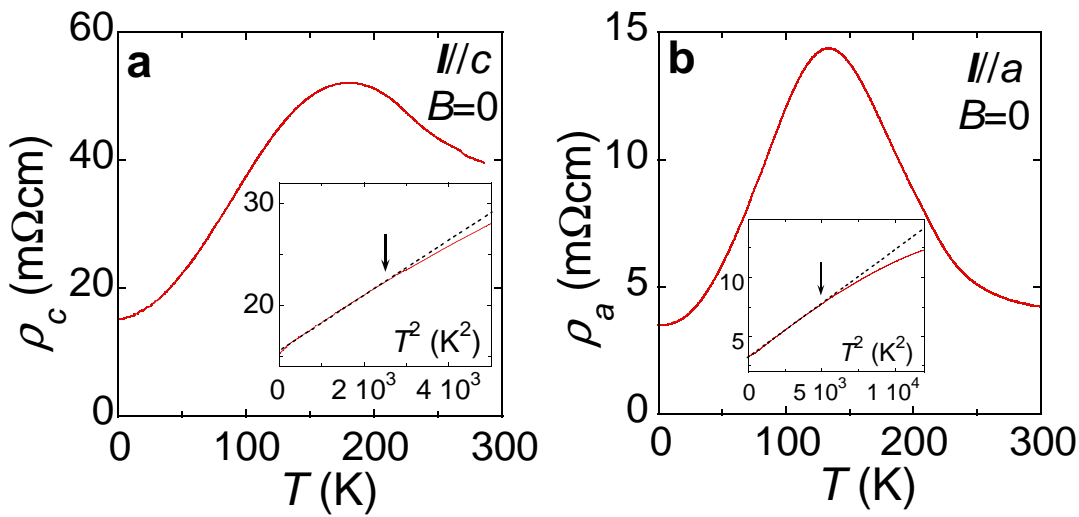}
\caption{{\bf a)} $\rho_{c}(T)$ for the same crystal as in Fig.
2a); the inset shows $\rho_{c}(T)$ versus $T^{2}$ below $T=70$K.
The arrow indicates the temperature $T_{\rm coh}^{c} = 50(5)$K at
which $\rho_{c}(T)$ first deviates from $T^{2}$. The kink in
$\rho_{c}(T)$ at low $T$ is due to the Pr ordering at $T=T_{\rm
N}$. {\bf b)} $\rho_{a}(T)$ for the same crystal as in Fig. 2b);
in the inset, $\rho_{a}(T)$ plotted versus $T^{2}$ below $T=110$K.
Here $T_{\rm coh}^{a}=70(5)$K (again indicated by an arrow).}
\label{Tcoh}
\end{figure}

\section{Disorder-induced localization in Pr124}

\begin{figure}[t]
\centering
\includegraphics[height=10.4cm, clip]{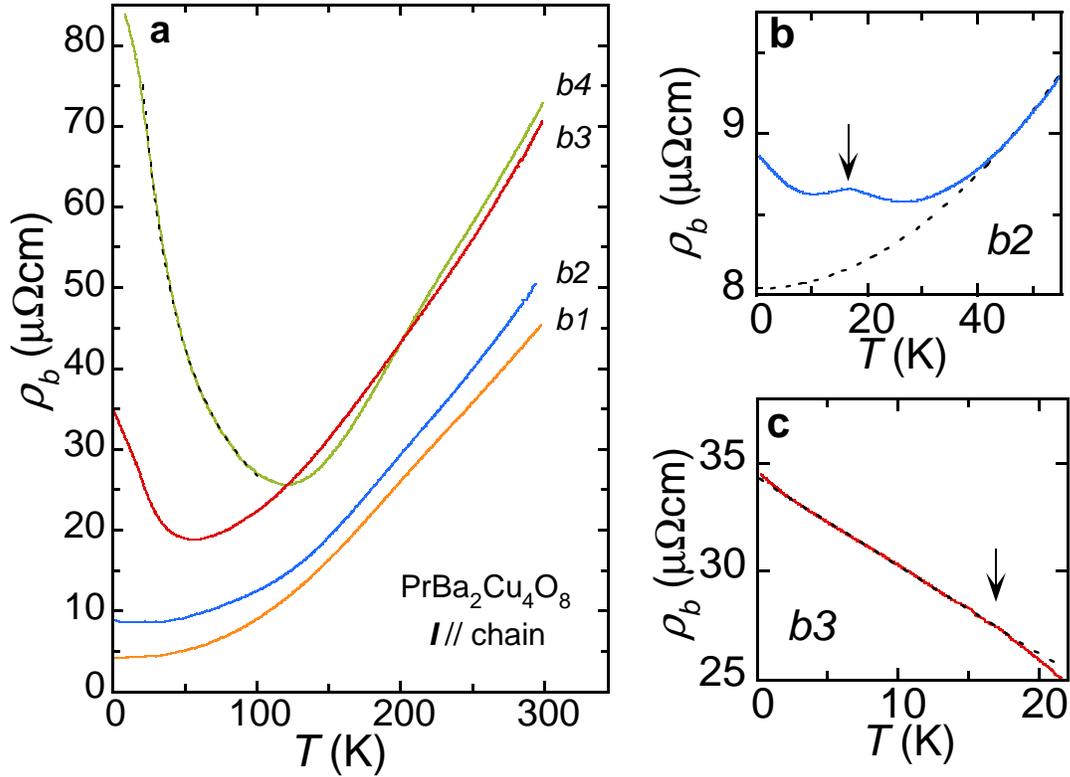}
\caption{{\bf a)} Zero-field $\rho_{b}(T)$ data for four single
crystals with different levels of disorder. The dashed line
overlaying the data of sample $b4$ is a fit to $\rho_{b}(T)$ =
$AT^{-2/3}$. {\bf b)} Blow-up of the low-$T$ $\rho_{b}(T)$ data
for sample $b2$, on the boundary between metallicity and
localization. The dashed line is an extrapolation of the metallic
$T^{2}$ dependence between 45K and 60K.
The arrow indicates the antiferromagnetic ordering temperature
$T_{N}$ of the Pr ions. {\bf c)} Blow-up of the low-$T$
$\rho_{b}(T)$ data for sample $b3$, highlighting the anomalous
$-T$ behaviour below $T_{\rm N}$ (again marked by an arrow).}
\label{Dis}
\end{figure}

We now turn our attention to the intrachain current response. Fig.
4a) shows $\rho_{b}(T)$ data for four needle-shaped samples $b1-4$
taken from the same growth batch. At high $T$, $\rho_{b}(T)$ is
metallic and quasi-linear, the slopes being similar in all four
samples suggesting uniform carrier concentration. Only $b1$
however remains metallic down to the lowest temperature studied
($T = 0.5$K). For $b1$, $\rho_{b}(T)= \rho_{0b}+AT^{\alpha}$ below
100K. The coefficient $\alpha = 2.3$ falls within the expected
range ($2 < \alpha < 3$) for a quasi-1D Fermi-liquid with dominant
electron-electron scattering~\cite{Oshiy78}. The other samples
have minima at $T = T_{\rm min}$ below which $\rho_{b}(T)$
gradually increases. Fig. 4b) shows a blow-up of the low-$T$
resistivity data for $b2$, located right at the boundary between
the metallic and non-metallic behaviour. The kink in $\rho_{b}(T)$
at $T_{N} = 17$K indicated by an arrow coincides with the
antiferromagnetic ordering of the Pr ions~\cite{Li00}. In $b3$ and
$b4$, this kink manifests itself as a change of slope.
Intriguingly, in {\it all} crystals that exhibit metallic
behaviour down to low $T$ (i.e. those with low residual
resistivities, including $b1$), $\rho_{b}(T)$ appears unaffected
by the Pr ordering.

Several mechanisms for the low $T$ insulating behaviour, including
variable range hopping~\cite{Mott}, Kondo spin
scattering~\cite{Kondo}, the charge Kondo
effect~\cite{Matsushita}, the ln(1/$T$) dependence observed in
underdoped 2D cuprates~\cite{Ando95} and the exponential behaviour
expected for a Mott insulator~\cite{GiamaBook}, were considered
but found to be incompatible with $\rho_b$($T$), both above and
below $T_N$. Whilst Fe concentrations below 100ppm can give rise
to resistivity upturns~\cite{Monod67}, we note here that the
in-chain magnetoresistance behaviour of all four of these crystals
is inconsistent with  Kondo scattering~\cite{NardaUnp}. Moreover
Mg was the only element whose concentrations were found to scale
with the size of the resistivity upturns; Mg$_{b2}: $Mg$_{b3}:
$Mg$_{b4}=3500:4200:4700$ (ppm). We therefore conclude that
magnetic impurities were not responsible for the upturns in
$\rho_{b}(T)$.

The large increase in $\rho_b$($T$) of $b3$ and $b4$ below $T_{\rm
min}$ argues against weak localization, whilst the very gradual
nature of the upturn in $\rho_b$($T$) suggests a lack of charge
ordering in Pr124, most probably due to the stabilizing presence
of the CuO$_2$ planes. The form of $\rho_b$($T$) for $T_{\rm N} <
T < T_{\rm min}$ is best represented by a power law, (e.g. dashed
line in Fig. 4a) for $b4$ where $\rho_{b}(T)$ varies as $T^{
-2/3}$). Below $T_{\rm N}$ however, the $T$-dependence of
$\rho_{b}(T)$ changes abruptly in all insulating samples. As
illustrated in Fig. 4c) for sample $b3$ for example, $\rho_b$($T$)
increases approximately linearly with decreasing $T$ over a decade
in temperature between 0.8K and 17K. The N\'{e}el ordering of the
Pr ions thus splits the insulating behaviour into two disparate
forms, one above $T_{\rm N}$ which is divergent, and one below
which is not, thereby making it difficult to identify the
intrinsic transport behaviour in the insulating state.

For a quasi-1D metal, the residual resistivity $\rho_{0b}$ is
independent of carrier density. Hence $\rho_{0b}$ can be used to
obtain a direct measure of the intrachain mean-free-path $\ell$.
Given that $b2$ lies on the threshold between metallicity and
localization, we can therefore extract an upper bound for the
nominal mean-free-path $\ell_{\rm cr}$ at the metal/insulator
boundary by extrapolating $\rho_{b}(T)$ of $b2$ from high $T$ down
to $T = 0$K. The dashed line in Fig. 4b) is an extrapolation of
the metallic $T^{2}$ dependence between 45K and 60K. From this we
obtain, $\rho_{0b} \leq 8(1)\mu$cm, giving $\ell_{\rm cr} \geq$
215(25)$\AA$. (Note that there are 2 chains per unit cell in
Pr124, and so $\rho_{0b}$ = $\pi ac \hbar/4e^{2}\ell$, where $a$
(= 3.88$\AA$) and $c$ (= 13.6$\AA$) are the $a$- and $c$-axis
lattice constants respectively.) Since the $b$-axis lattice
constant $b = 3.90 \AA$, this is equivalent to more than 50 unit
cells. Finally, taking estimates of the Fermi wave vector $k_{F}$
(= 0.2 $\AA^{-1}$) from angle-resolved photoemission spectroscopy
(ARPES)~\cite{Mizo00}, we arrive at $k_{F}\ell_{\rm cr} \geq
45(5)$ at the localization threshold. It is important to stress
that independent estimates of $\ell_{\rm cr}$ from transverse
interchain magnetoresistance measurements {\it that are
insensitive to any uncertainties in the crystal dimensions and
contact configurations} agree very well with the value extracted
from $\rho_{0b}$ of sample $b2$, thus supporting the convention
for voltage markers adopted here.

Localization at such large values of $k_{F}\ell$ is unprecedented.
In 3D Fermi-liquids for example, coherent (Bloch) electron motion
is destroyed once $k_{F}\ell < 1$, corresponding to a
mean-free-path shorter than the de Broglie wavelength. In the
normal state of the 2D cuprates, the localization threshold occurs
for $k_{F}\ell < 10$~\cite{Ando95}, though its origin is as yet
unknown. To our knowledge there have been no corresponding
experimental studies of the localization threshold in quasi-1D
systems. Extensive theoretical studies however have predicted that
an insulating phase develops in a {\it strictly} 1D system for a
vanishingly small amount of random impurities~\cite{MottTwose61,
Ishii73, GiamarchiSchulz88, KaneFisher92}. The tendency towards
localization in sample $b2$ at $k_{F}\ell_{\rm cr} \geq 45(5)$
suggests therefore that a fundamental change in the dimensionality
of the electronic system has occurred and indeed, direct
comparison of the relevant energy scales confirms this to be the
case.

\section{Discussion}

According to ARPES~\cite{Mizo00}, the Fermi velocity within the
chains $v_{F} = 2.5\times10^{5}$ ms$^{-1}$. Thus the intrachain
scattering rate at the localization threshold $\hbar/\tau_{\rm cr}
= \hbar v_{F}/\ell_{\rm cr} \leq 80(10)$K. As discussed above,
$2t_{\perp}^{a,c}$ (as determined from $B_{\rm cr}^{a,c}$) is also
proportional to $v_{F}$. Thus direct comparison of the three
energy scales ($2t_{\perp}^{a}$, $2t_{\perp}^{c}$ and
$\hbar/\tau_{\rm cr}$) is in fact independent of the value of
$v_{F}$. It is clear therefore that at the localization threshold,
$\hbar/\tau$ is comparable to both orthogonal hopping energies. In
this circumstance, intrachain scattering becomes sufficient to
block coherent wave propagation between chains, making them
decoupled electronically; the presence of arbitrary disorder
within the (effectively isolated) chains then leads immediately to
localization at the lowest temperature, as
predicted~\cite{MottTwose61, Ishii73, GiamarchiSchulz88,
KaneFisher92}.

As mentioned in the previous section, the form of $\rho_{b}(T)$
for $T_{\rm N}<T<T_{\rm min}$ is best represented by a power law,
(e.g. dashed line in Fig. 4a) for $b4$ where $\rho_{b}$ varies as
$T^{-2/3}$). Whilst this is consistent with predictions for a
Luttinger liquid~\cite{GiamaBook}, it should be stressed that the
$T$-range is too limited to make any concrete claims.
Nevertheless, recent optical~\cite{Take00} and ARPES~\cite{Mizo02}
results do support the emergence of Luttinger-liquid behaviour in
Pr124, either in pure Pr124 at high $T$ (i.e. when $k_{\rm
B}T>2t_{\perp}^{a,c})$ or in Zn-doped Pr124 at low $T$ (presumably
when $\hbar/\tau>2t_{\perp}^{a,c}$, though there no quantitative
estimates of the energy scales were made). From both these
studies, the Luttinger parameter $K_{\rho}$ is estimated to be
$\sim 0.24$, a value compatible~\cite{GiamaBook} with the
observation of a quasi-linear $T$-depenendence in $\rho_{b}$($T$)
at high $T$, if one assumes a carrier concentration close to, but
not quite, 1/4-filling. Significantly the resistivity data do not
exhibit the exponential behaviour expected for a 1D Mott insulator
at low $T$~\cite{GiamaBook}. The small value of $K_{\rho}$
suggests that this is due to deviations from commensurability,
rather than to the absence of strong interactions.

\section{Conclusions}

In this article, we have reported supporting evidence for the
first experimental realization of disorder-induced
one-dimensionality in a 3D solid at low temperatures. This has
only been made possible in Pr124 due to the extremely small values
of $2t_{\perp}$ ($\leq70$K) in {\it both} perpendicular
directions. In the quasi-1D organic conductors, such 3D to 1D
crossover phenomena are extremely difficult to induce (at low $T$)
due to the large coupling in the second direction (in
(TMTSF)$_2$PF$_6$ for example, $2t_{\perp}^{b} \sim
600$K~\cite{Vesco98}). This makes Pr124 a rather unique bulk
system with which to study physical phenomena on the boundary
between Fermi-liquid and Luttinger-liquid ground states that
complements existing work on chiral edge states in 2D
heterostructures~\cite{Milliken96}. Whilst it appears that
disorder allows the manifestation of one-dimensionality in Pr124
at low $T$, it is, by its very nature, one displaying localized
behaviour. Nevertheless, our results highlight one possible route
to a delocalised Luttinger-liquid at low $T$, in ultra-pure Pr124
under a tilted (pulsed) magnetic field with components in the $ac$
plane larger than the two crossover fields $B_{\rm cr}^{a,c}$.
Should such field-induced confinement also lead to an insulating
state, then the paradigm that all states are localized in real 1D
systems will gain yet more empirical confirmation. Finally, let us
remark on the observation that the dimensional crossovers in Pr124
appear to occur once the strength of a particular perturbation,
e.g. temperature $T$, magnetic field $B$, or scattering rate
$1/\tau$ , exceeds $2t_{\perp}$. This contrasts markedly with
recent studies on quasi-2D conductors where evidence for
interlayer coherence is observed despite the fact that both
$k_{\rm B}T$ and $\hbar/\tau \gg 2t_{\perp}$~\cite{Single02}.
Clearly the phenomenon of interlayer or interchain decoherence in
anisotropic metals is still poorly understood and it is only
through further systematic studies that this important issue can
be resolved.

\ack We acknowledge S. Andergassen, C. Capan, A. Carrington, R.
Claessen, T. Enss, T. Giamarchi, S. M. Hayden, S. Kivelson, N.
Shannon, I. Terasaki, V. Tripathi and J. A. Wilson for their
helpful comments. We also thank K. Nozawa and J. Hart for
technical assistance. This work was supported by the EPSRC (UK).
SH was financially supported by the Leverhulme Fellowship (UK).
The work at the NHMFL was supported by the National Science
Foundation and DOE Office of Science.

\appendix

\section{}

\begin{figure}[h]
\centering
\includegraphics[height=4cm,clip]{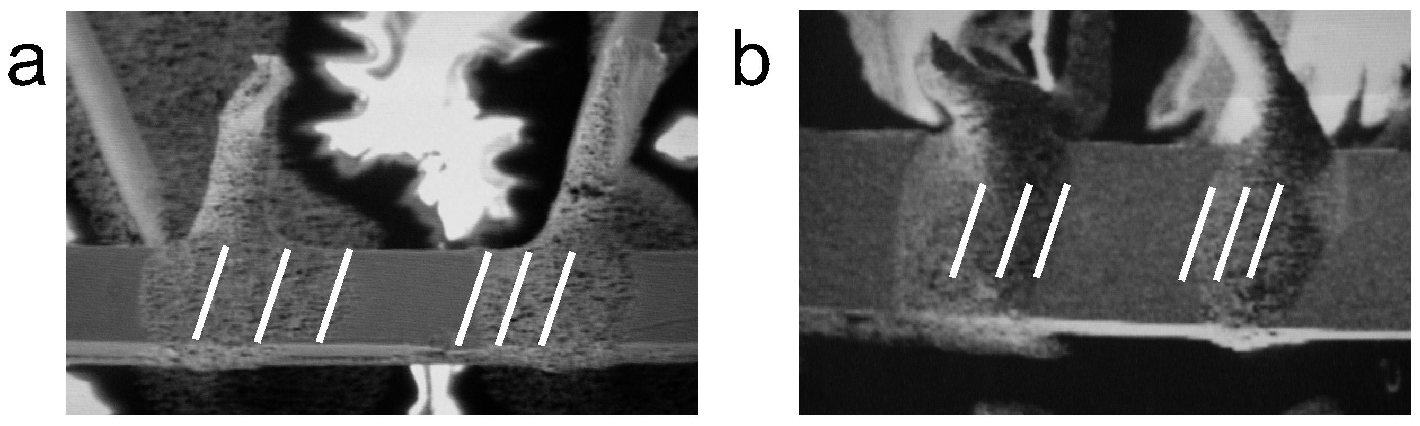}
\caption{{\bf a)} SEM picture of the pair of voltage contacts for
sample $b2$. The white lines represent different markers used to
determine the distance between contacts. {\bf b)} A similar
picture for sample $b3$. Using the markers at the mid-point of the
Au wires, the distances between voltage pads are 203$\mu$m and
195$\mu$m for samples $b2$ and $b3$ respectively.}
\label{Appendix}
\end{figure}

We address here the issue of the determination of absolute values
of the chain resistivity $\rho_{b}$. This requires an accurate
measurement of crystal dimensions and of distances between
contacts. With a high power optical microscope crystal dimensions
can be known to a reasonable degree of accuracy, 20-25\%, and with
an SEM, significantly better than that. The chief source of
systematic error is the large size of the electrical (voltage)
contacts relative to their separation. In order to ensure uniaxial
current flow in these highly anisotropic crystals it is necessary
to coat conductive paint across the entire sample in both
directions perpendicular to the current flow. This invariably
leads to large contact pads relative to the sample dimensions.
Moreover, it is not obvious which criterion one should use for
specifying the distance between the voltage contacts in
calculating the resistivity values. It depends to a large extent
on the contact resistance of each pad and where within the pad the
best electrical contact is made. A convention needs to be adopted
by associating a marker with each contact pad. As an example of
the importance of the marker scheme, Fig. \ref{Appendix} shows two
SEM pictures of the voltage contacts for samples $b2$ and $b3$.

Choosing the marker for the position of each contact at the Au
wire mid-point, at the edge of the Ag paste or at the intermediate
point between the two, leads to a factor of two or three
difference in the measured distance between contacts, and
therefore in the absolute value of $\rho_{b}$. We believe that the
discrepancies between values reported here and those reported
previously \cite{Hussey02, McBrien02, Nakada01} arise from
different conventions adopted to measure the distance between
voltage electrodes. In this study, we have used the distance
between the mid-point of the Au wires to calculate $\rho_{b}$.

\end{document}